\documentclass[aps,pra,twocolumn,showpacs,floatfix]{revtex4}

\usepackage{graphicx,amsmath,amssymb,xspace,epsfig,wasysym}

\let\oldsqrt\sqrt
\def\sqrt{\mathpalette\DHLhksqrt}
\def\DHLhksqrt#1#2{%
\setbox0=\hbox{$#1\oldsqrt{#2\,}$}\dimen0=\ht0
\advance\dimen0-0.2\ht0
\setbox2=\hbox{\vrule height\ht0 depth -\dimen0}%
{\box0\lower0.4pt\box2}}

\newcommand{\bseq}{\begin{eqnarray}}
\newcommand{\eseq}{\end{eqnarray}}
\newcommand{\half}{\mbox{$\frac{1}{2}$}}
\newcommand{\sfrac}[2]{\mbox{$\frac{#1}{#2}$}}
\newcommand{\mb}[1]{\mathbf{#1}}
\newcommand{\bs}[1]{\boldsymbol{#1}}
\newcommand{\uu}{{\uparrow\uparrow}}
\newcommand{\ud}{{\uparrow\downarrow}}
\newcommand{\du}{{\downarrow\uparrow}}
\newcommand{\dd}{{\downarrow\downarrow}}
\newcommand{\dis}[1]{$\displaystyle #1 $}

\begin{document}

\title{Density Functional of a Two-Dimensional Gas of Dipolar Atoms:\\ 
Thomas-Fermi-Dirac Treatment}

\author{Bess Fang}
\author{Berthold-Georg Englert}
\affiliation{%
  Centre for Quantum Technologies, National University of
  Singapore, 3 Science Drive 2, Singapore 117543, Singapore\\
  Department of Physics, Faculty of Science, National
  University of Singapore, 2 Science Drive 3, Singapore 117542, Singapore}

\date{6 August 2010}

\begin{abstract}
We derive the density functional for the ground-state energy of a
two-dimensional, spin-polarized gas of neutral fermionic atoms with
magnetic-dipole interaction, in the Thomas-Fermi-Dirac approximation.  
For many atoms in a harmonic trap, we give analytical solutions for the
single-particle spatial density and the ground-state energy, in dependence on
the interaction strength, and we discuss the weak-interaction limit that is
relevant for experiments.
We then lift the restriction of full spin polarization and account for a
time-independent inhomogeneous external magnetic field.
The field strength necessary to ensure full spin polarization is derived.
\end{abstract}

\pacs{31.15.E-, 71.10.Ca}

\maketitle

\section{Introduction}

Ultracold atomic gases provide highly controllable systems for the study of
condensed-matter phenomena \cite{bloch}.  With the ongoing experimental
efforts in ultracold gases of neutral Fermi atoms \cite{demarco} and the
possibility of a genuinely two-dimensional (2D) confinement
\cite{modugno,jochim,guenther,martinyanov,dyke}, it is now possible to acquire
data about dilute 2D degenerate Fermi gases of neutral atoms under adjustable
laboratory conditions \cite{froehlich}.  It is hoped that this will advance our
understanding of various 2D phenomena, such as high-$T_c$ superconduction
\cite{lee,leggett}, effective massless Dirac fermions \cite{zhu,LeeKL}, the
BEC-BKT cross over \cite{petrov,zhang}, and others.

While experiments of this kind will involve periodic potentials of various
geometries, all experiments with ultracold atoms in a 2D configuration 
will make use of a confining potential.  
Before investigating aspects of specific periodic potentials, one can examine 
the behavior of the degenerate gas in the 2D trapping potential.  
We study this situation with the help of density functionals, and arrive at
detailed predictions for the case of an isotropic harmonic trapping potential.  

Density functional theory (DFT), which has its historical roots in the
Thomas-Fermi model for atoms \cite{thomas,fermi}, was first formulated for the
inhomogeneous electron gas \cite{hohenberg}, with immediate applications to
atoms, molecules, and solids; see Ref.~\cite{dreizler}, for example.
DFT can equally well be used for studying other physical systems, such as
dilute gases of neutral fermionic atoms under the influence of a confining
external potential.
While the DFT formalism can be based on both the spatial \cite{hohenberg} 
and the momental density \cite{henderson}, the spatial-density version gives a
more natural description in the case of a position-dependent interaction, 
such as the magnetic dipole interaction.  
We derive the density functionals for spin-polarized fermions with magnetic
dipole interaction, confined in a 2D harmonic potential, and investigate the
ground-state density and energy of the system.

The article is organized as follows.  
Section \ref{sec:3D} summarizes earlier investigations in three dimensions
(3D).  
In Sec.\ \ref{sec:3to2}, we discuss how to appropriately reduce the
dimensionality.  
Various 2D density functionals are derived in Sec.\ \ref{sec:fnal}.  
A discussion about the scaling behavior of these functionals is then given in
Sec.\ \ref{sec:virial}.  
Section \ref{sec:result} presents the analytic results of the ground-state
density and energy, and discusses the weak interaction limit.  
Section \ref{sec:SDM} extends the formalism to accommodate the spin dependence
in an inhomogeneous magnetic field.   
We conclude with a summary and a brief outline of prospective work.

\section{The 3D case}\label{sec:3D}
It is expedient to recall some basic relations that were earlier established
in 3D, mainly collected from Refs.\ \cite{englert1,englert2,cinal,goral}.

\subsection{Single-particle density and density matrix}
The spatial one-particle density matrix $n^{(1)}({\mathbf r'};{\mathbf r''})$
and the associated one-particle Wigner function 
$\nu ({\mathbf r},{\mathbf p})$ are related by
\begin{equation}\label{Eqn:n1}
n^{(1)}({\mathbf r'};{\mathbf r''}) 
= \int \frac{(d{\mathbf p})}{(2 \pi \hbar)^3} \,
\nu \bigl(\tfrac{1}{2}(\mathbf{r}' +\mathbf{r}''), 
\mathbf{p} \bigr)\,e^{i\mathbf{p} \cdot (\mathbf{r}'-\mathbf{r}'')/\hbar}\, ,
\end{equation}
with $(d\mathbf{p}) \equiv dp_xdp_ydp_z$ denoting the volume element in the
momentum space.  
The spatial and momental one-particle densities are obtained by integrating 
$\nu({\mathbf r},{\mathbf p})$ over the other variable,
\begin{eqnarray}
 n(\mathbf{r})  \mathrel{\equiv} n^{(1)}(\mathbf{r};\mathbf{r})
& = & \int \frac{(d\mathbf{p})}{(2 \pi \hbar)^3} 
      \,\nu(\mathbf{r},\mathbf{p})\, , \nonumber \\
\rho(\mathbf{p}) & = & \int \frac{(d\mathbf{r})}{(2 \pi \hbar)^3} 
\,\nu(\mathbf{r},\mathbf{p})\ .\label{Eqn:n}
\end{eqnarray}
Note that both densities are normalized to the total number of particles $N$,
\begin{equation} \label{Eqn:norm}
N = \int (d\mathbf{r})~n({\mathbf r}) = \int (d\mathbf{p})~\rho(\mathbf{p})\,.
\end{equation}

\subsection{Density functionals for energy}
For a system of spin-polarized fermions in an isotropic harmonic trap, the
potential energy is given by
\begin{equation}\label{Eqn:Etrap}
E_{\mathrm{trap}}[n] = \int (d\mathbf{r}) \,\frac{1}{2} M \omega^2 r^2
\,n(\mathbf{r})\, ,
\end{equation}
where $M$ and $\omega$ are the mass of an individual atom and the trap
frequency, respectively, $r=|\mathbf{r}|$ is the length of the position vector
$\mathbf{r}$, and the kinetic energy is
\begin{equation}
E_{\mathrm{kin}} = \int (d\mathbf{p}) \,\frac{p^2}{2M} \,\rho(\mathbf{p})\, .  
\end{equation}
Both $E_{\mathrm{trap}}$ and $E_\mathrm{kin}$ are sums over single-particle
contributions.  

The interaction energy, $E_\mathrm{dd}$, which is a sum over particle-pair
contributions, is evaluated using the diagonal part of the two-particle
density matrix
$n^{(2)}(\mathbf{r}_1',\mathbf{r}_2';\mathbf{r}_1'',\mathbf{r}_2'')$,
\begin{equation}
E_\mathrm{dd} = \frac{1}{2} \int
(d\mathbf{r}')(d\mathbf{r}'')\,V_\mathrm{dd}(\mathbf{r}'-\mathbf{r}'')
\,n^{(2)}(\mathbf{r}',\mathbf{r}'';\mathbf{r}',\mathbf{r}'')\, 
\end{equation}
with the magnetic dipolar interaction potential
\begin{equation}\label{Eqn:Vdd}
V_\mathrm{dd}(\mathbf{r}) = \frac{\mu_0}{4\pi} \biggl[ \frac{\mu^2}{r^3} -
3 \frac{(\boldsymbol{\mu}\cdot\mathbf{r})^2}{r^5} -
\frac{8\pi}{3}\mu^2\delta(\mathbf{r}) \biggr]\,,
\end{equation}
where $\boldsymbol{\mu}$ and $\mu$ are the magnetic dipole moment and its
magnitude of an individual atom.  
The contact term in $V_{\mathrm{dd}}(\mathbf{r})$ is necessary to ensure that the
magnetic field generated by the point dipole is divergence-free.

\subsection{TFD approximation}
In the spirit of the approach that was pioneered by Thomas,
Fermi, and Dirac (TFD), a two-fold semiclassical
approximation is employed here.  First, $n^{(2)}$ is replaced by products of
$n^{(1)}$ factors (due to Dirac~\cite{dirac}) according to
\begin{eqnarray}\label{eqn:n2Dirac}
    n^{(2)}(\mathbf{r}_1',\mathbf{r}_2';\mathbf{r}_1'',\mathbf{r}_2'') & = &
    n^{(1)}(\mathbf{r}_1'; \mathbf{r}_1'')n^{(1)}(\mathbf{r}_2';
    \mathbf{r}_2'') \nonumber \\ & & \mbox{}
- n^{(1)}(\mathbf{r}_1'; \mathbf{r}_2')n^{(1)}(\mathbf{r}_2'';\mathbf{r}_1'') .
\end{eqnarray}
This splitting corresponds to the direct and exchange terms when
evaluating the interaction energy, $E_\mathrm{dd}$.  
Note that this expression is only valid if the system is spin-polarized.  
Otherwise, a multiplicative
constant preceding the second term is needed to account for the
spin-multiplicity.  

Second, the Wigner function is a uniform sphere of a
finite size (due to Thomas~\cite{thomas} and Fermi~\cite{fermi})
\begin{equation}\label{Eqn:nu}
\nu(\mathbf{r},\mathbf{p}) 
= \eta\bigl(\hbar[6\pi^2n(\mathbf{r})]^{1/3} - p\bigr)\,,
\end{equation}
where $\eta(~)$ is the Heaviside unit step function.  This applies when
functionals of the spatial density $n(\mb{r})$ are considered. 
For functionals of the momental density $\rho(\mb{p})$, one has to use
$\nu(\mathbf{r},\mathbf{p}) = \eta\bigl(t(\mathbf{p})-V(\mathbf{r})\bigr)$
where $V(\mathbf{r})$ is the external potential and $t(\mathbf{p})$ is
determined by $\rho(\mb{p})$ through Eq.~(\ref{Eqn:n}).
In the case of an isotropic harmonic potential, $V(\mb{r})\propto r^2$, this
is
\begin{equation}
\nu(\mathbf{r},\mathbf{p}) = \eta\bigl(\hbar\!\left[6\pi^2
  \rho(\mathbf{p})\right]^{1/3} - r\bigr)\,,  
\end{equation}
visibly the analog of Eq.~(\ref{Eqn:nu}).

This yields the familiar density functional of the kinetic energy, 
\begin{equation}\label{Eqn:Ekinn}
E_\mathrm{kin} [n] = \int (d\mathbf{r}) \,\frac{\hbar^2}{M} \,\frac{1}{20\pi^2}
\left[ 6\pi^2 n(\mathbf{r})\right]^{5/3}\, .
\end{equation}
Since the contributions of the contact term to the direct and to the exchange
energy cancel each other in the fully spin-polarized situation under
consideration, and the remaining exchange energy vanishes under the average
over the solid angle associated with the relative distance, the density
functional of the interaction energy,
\begin{equation}
  \label{Eqn:Vddbar1}
  E_\mathrm{dd}[n] = \frac{1}{2} \int \!(d\mathbf{r}) (d\mathbf{r}') 
\,n(\mathbf{r})\,\overline{V}_\mathrm{dd}(\mathbf{r}-\mathbf{r}') 
\,n(\mathbf{r}')\, ,
\end{equation}
is characterized by an effective potential $\overline{V}_\mathrm{dd}$,
\begin{equation}  \label{Eqn:Vddbar2}
\overline{V}_\mathrm{dd} (\mathbf{r}) = \frac{\mu_0}{4\pi} \left[
  \frac{\mu^2}{r^3} -
  3\frac{(\boldsymbol{\mu}\cdot\mathbf{r})^2}{r^5} \right]\, .  
\end{equation}

\subsection{Ground-state energy and density}
Thus, the functional for the total energy of the ground state in the TFD
approximation is given by the sum of the three terms in 
Eqs.~(\ref{Eqn:Etrap}), (\ref{Eqn:Ekinn}), and (\ref{Eqn:Vddbar1}),
\begin{eqnarray}\label{Eqn:ETFDn}
E_\mathrm{TFD}[n] & = & \int \!(d\mathbf{r}) \frac{\hbar^2}{M}\, 
\frac{1}{20\pi^2}
  [ 6\pi^2 n(\mathbf{r})]^{5/3} \nonumber\\
& &  + \int \!(d\mathbf{r}) \frac{1}{2} M\omega^2r^2
  n(\mathbf{r}) \nonumber\\
& & + \frac{1}{2} \int \!(d\mathbf{r}) (d\mathbf{r}')\,
n(\mathbf{r}) \,\overline{V}_\mathrm{dd}(\mathbf{r}-\mathbf{r}')
\,n(\mathbf{r}')\,.\qquad
\end{eqnarray}
Upon applying the variational principle, 
we find that the density that minimizes 
$E_\mathrm{TFD}$ must obey the integral equation
\begin{eqnarray}\label{Eqn:VP}
\frac{\hbar^2}{2M} \left[ 6\pi^2n(\mathbf{r}) \right]^{2/3} + \frac{1}{2}
M\omega^2r^2 \hspace*{2em}&&\nonumber \\
\mbox{}+ 
\int \!(d\mathbf{r}')\,\overline{V}_{\mathrm{dd}}(\mathbf{r}-\mathbf{r}') 
\,n(\mathbf{r}')& =& \frac{1}{2} M\omega^2 R^2\, ,\qquad
\end{eqnarray}
where $\frac{1}{2}M\omega^2R^2$ is a convenient way of parameterizing the
Lagrange multiplier for the normalization constraint of Eq.~(\ref{Eqn:norm}).

\section{From 3D to 2D}\label{sec:3to2}
The form of the density functional in Eq.~(\ref{Eqn:ETFDn})
gives no explicit indication of its dependence on the spatial dimension.  It is
thus necessary to re-derive the density functionals in 2D, with some
suitable assumptions about the Wigner function.

\subsection{A possible Wigner function}
When the trapping potential in the $z$-direction is harmonic and sufficiently
stiff, as is the typical situation in an actual experiment,
the system will remain in the ground state in this direction, and this
gives rise to a factorizable Gaussian dependence in $z$ and $p_z$ in the
Wigner function,
\begin{equation}\label{Eqn:nu32}
\nu(\mathbf{r},\mathbf{p}) = \nu_\perp (\mathbf{r}_\perp, \mathbf{p}_\perp)\, 2
\exp{\left( -\frac{z^2}{l_z^2} - \frac{p_z^2l_z^2}{\hbar^2} \right)} ,
\end{equation}
where $l_z = \sqrt{\hbar/(M\omega_z)}$ is the harmonic oscillator length scale
in the $z$-direction, and the subscript `$_\perp$' indicates that these
various quantities live in the transverse $xy$-plane.  
Here, $\omega_z$ is a finite but large frequency and,
in order to achieve a 2D geometry,  we require that
${\hbar \omega_z \gg k_B T}$ for the situation of ultracold atoms that we have
in mind, although we take the limit
$\omega_z \rightarrow \infty$ for mathematical convenience whenever possible.

\subsection{Densities in 2D}
In analogy with the densities defined in 3D, Eqs.~(\ref{Eqn:n1}) and
(\ref{Eqn:n}), the densities in 2D are given by,
\begin{eqnarray}
  n^{(1)}_\perp({\mathbf r}_\perp';{\mathbf r}_\perp'') 
& = & \int \!\frac{(d{\mathbf p}_\perp)}{(2 \pi \hbar)^2} \,\nu_\perp 
\bigl(\tfrac{\mathbf{r}_\perp'+\mathbf{r}_\perp''\rule[-3pt]{0pt}{2pt}}{2}, 
 {\mathbf p}_\perp\bigr)\,
e^{i\mathbf{p}_\perp \cdot (\mathbf{r}_\perp'-\mathbf{r}_\perp'')/\hbar}\,,\nonumber \\
  n_\perp(\mathbf{r}_\perp) & = & 
\int \!\frac{(d\mathbf{p}_\perp)}{(2 \pi \hbar)^2}
  \,\nu_\perp(\mathbf{r}_\perp,\mathbf{p}_\perp)\,,\nonumber \\
  \rho_\perp(\mathbf{p}_\perp) & = & 
\int \!\frac{(d\mathbf{r}_\perp)}{(2 \pi \hbar)^2} 
\,\nu_\perp(\mathbf{r}_\perp,\mathbf{p}_\perp)\,.
\label{Eqn:rho2D}
\end{eqnarray}
With the decomposition of the Wigner function in
Eq.~(\ref{Eqn:nu32}), we find that the densities in 2D and those in 3D are
related in the following manner:
\begin{eqnarray}
n^{(1)}(\mathbf{r}';\mathbf{r}'') & = &
n^{(1)}_\perp(\mathbf{r}_\perp';\mathbf{r}_\perp'')
\,\frac{1}{l_z\sqrt{\pi}} \exp{\left( -\frac{4z_+^2+z_-^2}{4l_z^2} \right)},
\nonumber \\ 
n(\mathbf{r}) & = & n_\perp(\mathbf{r}_\perp) 
\,\frac{1}{l_z\sqrt{\pi}} \exp{\left( -\frac{z^2}{l_z^2} \right)},\nonumber \\
\rho(\mathbf{p}) & = & \rho_\perp(\mathbf{p}_\perp)
\,\frac{l_z}{\hbar\sqrt{\pi}} \exp{\left( -\frac{p_z^2 l_z^2}{\hbar^2}
\right)},\label{Eqn:n32}
\end{eqnarray}
where $z_+=\frac{1}{2}(z'+z'')$, and $z_-=z'-z''$, 
such that the 2D densities are now normalized to the number of particles,
\begin{equation}
N = \int \!(d\mathbf{r}_\perp) \,n_\perp(\mathbf{r}_\perp) = \int
\!(d\mathbf{p}_\perp) \,\rho_\perp(\mathbf{p}_\perp)\, .\label{Eqn:norm2D} 
\end{equation}

\subsection{Various energy terms}
By integrating over $z$ and $p_z$, we immediately find the trap energy and the
kinetic energy in terms of the 2D densities,
\begin{eqnarray}
  E_\mathrm{trap} & = & \int \!(d\mathbf{r}_\perp) \,n_\perp(\mathbf{r}_\perp)
  \,\frac{1}{2}M\omega_\perp^2 r_\perp^2 + \frac{N}{4} \hbar \omega_z \,, 
\nonumber \\
  E_\mathrm{kin} & = & \int \!(d\mathbf{p}_\perp) \,\rho_\perp(\mathbf{p}_\perp)
  \,\frac{p_\perp^2}{2M} + \frac{N}{4} \hbar \omega_z \,,\label{Eqn:Ekin2}
\end{eqnarray}
where $\omega_\perp$ is the radial trap frequency in the $xy$-plane, assuming
isotropy.  Note that both expressions contain parts analogous to the
corresponding expressions in 3D and additive constants, which are the sum of
single-particle energies in the ground state of the harmonic trap of the
$z$-confinement.  Since these constants play no role in the dynamics of the
system, we renormalize the expressions, such that
\begin{eqnarray}\label{Eqn:EtkRN}
  E_\mathrm{trap} & = & \int \!(d\mathbf{r}_\perp) \,n_\perp(\mathbf{r}_\perp)
  \,\frac{1}{2}M\omega_\perp^2 r_\perp^2 \,,\nonumber \\
  E_\mathrm{kin} & = & \int \!(d\mathbf{p}_\perp) \,\rho_\perp(\mathbf{p}_\perp)
  \,\frac{p_\perp^2}{2M}\,,
\end{eqnarray}
which are now independent of $\omega_z$ and unaffected when the limit
$\omega_z \rightarrow \infty$ is taken.  

To investigate the  interaction energy, we employ the (2+1)D version of
Eq.~(\ref{eqn:n2Dirac}), 
\begin{eqnarray}\label{Eqn:n2split}
  n^{(2)}(\mathbf{r}',\mathbf{r}'';\mathbf{r}',\mathbf{r}'') & = &
  n(\mathbf{r}') \,n(\mathbf{r}'') - n^{(1)}(\mathbf{r}';\mathbf{r}'')
  \,n^{(1)}(\mathbf{r}'';\mathbf{r}') \nonumber \\
  & = & \frac{e^{-(4z_+^2+z_-^2)/(2 l_z^2)}}{l_z^2 \pi}  \Bigl(
  n_\perp(\mathbf{r}_\perp') \,n_\perp(\mathbf{r}_\perp'') \nonumber\\
  & & \mbox{} - n^{(1)}_\perp (\mathbf{r}_\perp';\mathbf{r}_\perp'')
  \,n^{(1)}_\perp(\mathbf{r}_\perp'';\mathbf{r}_\perp') \Bigr)\, ,
\end{eqnarray}
which corresponds to the splitting of the direct and exchange energies.  It is
clear from Eq.~(\ref{Eqn:n2split}) that the contact term in the interaction
potential, Eq.~(\ref{Eqn:Vdd}), enforces $\mathbf{r}'=\mathbf{r''}$ and thus
equal and opposite contributions from the direct and exchange energies, the
familiar situation when the system is spin-polarized.  It is then permissible
to drop the contact term, which amounts to replacing the original interaction
potential by the effective potential of Eq.~(\ref{Eqn:Vddbar2}), i.e.\
\begin{eqnarray}
  E_\mathrm{dd} & = & \frac{1}{2} \int
  \!(d\mathbf{r}')(d\mathbf{r}'')\,\overline{V}_{\rm
    dd}(\mathbf{r}'-\mathbf{r}'') \nonumber \\
& & \times \Bigl( n(\mathbf{r}') \,n(\mathbf{r}'') -
  n^{(1)}(\mathbf{r}';\mathbf{r}'') \,n^{(1)}(\mathbf{r}'';\mathbf{r}')
  \Bigr)\,.
\quad 
\end{eqnarray}
Since $\overline{V}_\mathrm{dd}(\mathbf{r}'-\mathbf{r}'')$ depends only on the
difference in the positions, we isolate the $z$-direction by identifying
$\boldsymbol{\rho} = (\mathbf{r}'-\mathbf{r}'')_\perp$, 
\begin{equation} \label{Eqn:vdd}
  \overline{V}_\mathrm{dd}(\mathbf{r}'-\mathbf{r}'') 
  = \frac{\mu_0\mu^2}{4\pi} \left[
    \frac{1}{(\rho^2+z_-^{~2})^{3/2}} -
    \frac{3z_-^{~2}}{(\rho^2+z_-^{~2})^{5/2}} \right],
\end{equation}
where we assume that the magnetic dipole moments of all fermions are polarized
in the $z$-direction, i.e.\ $\boldsymbol{\mu} = \mu\hat{\mathbf{e}}_z$.  

It should be noted that the replacement of $V_\mathrm{dd}(\mathbf{r})$ by
$\overline{V}_\mathrm{dd}(\mathbf{r})$ only takes place after the
approximation in Eq.~(\ref{Eqn:n2split}) is made. 
It may not be correct for a different approximation scheme, that is: when
going beyond Dirac's approximation in Eq.~(\ref{eqn:n2Dirac}).

In the limit of $\omega_z \rightarrow \infty$, the Gaussians of $z_\pm$
become Dirac delta functions, yielding
\begin{eqnarray}
E_\mathrm{dd} & = & 
\frac{1}{2} \int \!(d\mathbf{r}_\perp')(d\mathbf{r}_\perp'')\,
  \frac{\mu_0\mu^2}{4\pi} \frac{1}{|\mathbf{r}_\perp'-\mathbf{r}_\perp''|^3}\
  \nonumber\\ 
&  &  \times \Bigl( \!n_\perp\!(\mathbf{r}_\perp')
  n_\perp(\mathbf{r}_\perp'')-
  n_\perp^{(1)}(\mathbf{r}_\perp';\mathbf{r}_\perp'')
  n_\perp^{(1)}(\mathbf{r}_\perp'';\mathbf{r}_\perp') \Bigr)\,.  \nonumber\\
\end{eqnarray}
In hindsight, we recognize the result above as an immediate consequence of
having $\boldsymbol{\mu} \perp \mathbf{r}$, which forces their scalar product
in Eq.~(\ref{Eqn:Vddbar2}) to vanish, while the contact term does not
contribute for the reason discussed earlier.

\section{2D functionals}\label{sec:fnal}
From this section onwards, for notational convenience, we leave out all the
subscripts `$_\perp$'.  It is understood that all the densities refer to the
2D definition specified in Eqs.~(\ref{Eqn:rho2D}), and all the vectors reside
in the $xy$ plane.

\subsection{TFD: A brutally simple Wigner function}
In order to derive the density functionals in 2D, one first needs to complete
the TFD approximation started in the previous section and assume a ``brutally
simple Wigner function'' \cite{goral} analogous to Eq.~(\ref{Eqn:nu}), i.e.\
\begin{equation}\label{Eqn:nu2}
\nu(\mathbf{r},\mathbf{p}) = 
\eta\bigl( \hbar [4\pi n(\mathbf{r})]^{1/2} - p \bigr)\, , 
\end{equation}
where the power and prefactor of the density are determined by normalization.

\subsection{Density functionals}
Not surprisingly, the potential energy reads
\begin{equation}\label{Eqn:Etrap2D}
E_\mathrm{trap} [n] = \int \!(d\mathbf{r}) \,\frac{1}{2} M\omega^2 r^2
\,n(\mathbf{r})\, ,
\end{equation}
where we emphasize that $\omega$ is the radial trap frequency in the
$xy$-plane, assuming isotropy.  Upon using Eqs.~(\ref{Eqn:rho2D}),
(\ref{Eqn:EtkRN}), and (\ref{Eqn:nu2}), we find that
\begin{equation}\label{Eqn:Ekin2D}
E_\mathrm{kin} [n] = \int \!(d\mathbf{r}) \,\frac{\hbar^2}{M}\, \pi
\,n(\mathbf{r})^2\, .  
\end{equation}
The $n(\mathbf{r})^2$ dependence of this 2D functional can also be
obtained from dimensional analysis; similarly, dimensional analysis confirms
the  $n(\mathbf{r})^{5/3}$ dependence of the 3D functional 
in Eq.~(\ref{Eqn:Ekinn}), and the 1D functional for the kinetic energy has the
cube of the density; see Table~\ref{Tab:123D} below. 

The interaction energy, in particular, turns out to consist of two pieces
with different dependence on the one-particle density, namely
\begin{eqnarray}\label{Eqn:Edd2D}
  E_\mathrm{dd}^{\ } & = & E_\mathrm{dd}^{(1)} + E_\mathrm{dd}^{(2)}\, , \nonumber \\
  E_\mathrm{dd}^{(1)} [n] & = & \frac{\mu_0\mu^2}{4\pi} \int \!(d\mathbf{r})
  \frac{256}{45} \sqrt{\pi}\, n(\mathbf{r})^{5/2}\, , \nonumber\\
  E_\mathrm{dd}^{(2)} [n] & = & -\frac{\mu_0\mu^2}{4\pi} \pi \int \!(d\mathbf{r})
  \,n(\mathbf{r}) \sqrt{-\nabla^2} n(\mathbf{r})\, ,
\end{eqnarray}
where $\sqrt{-\nabla^2}$ is an integral operator that is given by
\begin{equation}\label{Eqn:RNL}
\sqrt{-\nabla^2} n(\mathbf{r}) = \int \!\frac{(d\mathbf{r}')}{(2\pi)^2}
(d\mathbf{k})\, k\,e^{-i\mathbf{k}\cdot(\mathbf{r}-\mathbf{r}')}
n(\mathbf{r}')\, .
\end{equation}
We report the details of deriving Eqs.~(\ref{Eqn:Edd2D}) in the Appendix.  

Note that the splitting of the interaction energy in Eqs.~(\ref{Eqn:Edd2D})
does not correspond to the direct and exchange energies as in the 3D case,
where, as we recall, the exchange energy exactly compensates for the
contribution of the contact term to the direct energy.  In 2D, both the direct
and exchange energies are infinite individually, and they can only be
considered together so that the total interaction energy is finite.  Both
contributions in Eqs.~(\ref{Eqn:Edd2D}) stem from the sum of the direct and
the exchange energy.

\subsection{Ground-state energy and density}
The total energy of the system in the TFD approximation is now given
by the sum of the various energy terms derived above,
\begin{eqnarray}
& & E_\mathrm{TFD}^\mathrm{(2D)} [n] 
= \int \!(d\mathbf{r}) \,\frac{\hbar^2}{M} \pi
\,n(\mathbf{r})^2 + \int \!(d\mathbf{r}) \,\frac{1}{2}M\omega^2 r^2
n(\mathbf{r}) \nonumber \\
 & &  \mbox{}+ \frac{\mu_0\mu^2}{4\pi} \int \!(d\mathbf{r})\biggl[
  \frac{256}{45} \sqrt{\pi} \, n(\mathbf{r})^{5/2} - \pi
  n(\mathbf{r}) \sqrt{-\nabla^2} n(\mathbf{r}) \biggr] \,.\nonumber\\
\end{eqnarray}
The density that minimizes the energy, constrained by the normalization
condition (\ref{Eqn:norm2D}), must then obey
\begin{eqnarray}\label{Eqn:VP2D}
& & \frac{2\hbar^2}{M} \pi n(\mathbf{r}) + \frac{1}{2}M\omega^2r^2 +
\frac{\mu_0\mu^2}{4\pi} \bigg[ \frac{128}{9} \sqrt{\pi}\, n(\mathbf{r})^{3/2}
\nonumber \\
  & & \hspace*{8em}
  - 2\pi \sqrt{-\nabla^2} n(\mathbf{r}) \bigg] = \frac{1}{2}M\omega^2R^2\,,
\qquad
\end{eqnarray}
where, as in Eq.~(\ref{Eqn:VP}), $\frac{1}{2}M\omega^2R^2$ is the chemical
potential.

A comparison between Eqs.~(\ref{Eqn:VP2D}) and (\ref{Eqn:VP}) shows
that the reduction of dimension does not yet provide any operational
simplification when it comes to solving for the spacial density, because of
the occurrence of the integral operator $\sqrt{-\nabla^2}$.
But one should not fail to notice that the interaction is now made up of two
contributions with different dependence on the density.  
As we will see in the next section, the integral term is rather unimportant in
certain parameter regimes of interest and can then be neglected.

\section{Virial theorem and Scaling}\label{sec:virial}

\subsection{Scaling transformation}
Let us consider scaling transformations that change both the length scale and
the number of particles,
\begin{equation}
n(\mathbf{r}) \rightarrow \lambda^{2+\alpha}n(\lambda\mathbf{r})\, ,\quad 
N\rightarrow \lambda^\alpha N\, .
\end{equation}
They are consistent with the normalization constraint,
Eq.~(\ref{Eqn:norm2D}), and affect the various terms of 
$E_\mathrm{TFD}^{\mathrm{(2D)}}$ in the following manner:
\begin{eqnarray}
E_\mathrm{kin} & \rightarrow & \lambda^{2+2\alpha}E_\mathrm{kin}\,, 
\nonumber \\
E_\mathrm{trap} & \rightarrow & \lambda^{-2+\alpha} E_\mathrm{trap}\,, 
\nonumber \\
E_\mathrm{dd}^{(1)} & \rightarrow & \lambda^{3+5\alpha/2}E_\mathrm{dd}^{(1)}\,, 
\nonumber \\
E_\mathrm{dd}^{(2)} & \rightarrow & \lambda^{3+2\alpha}E_\mathrm{dd}^{(2)}\,,
\end{eqnarray}
so that the total energy $E \mathrel{\equiv} E_\mathrm{TFD}^\mathrm{(2D)}$ 
transforms in accordance with
\begin{eqnarray} \label{Eqn:SumE}
E &=& E_{\mathrm{kin}} + E_{\mathrm{trap}} + E_{\mathrm{dd}}^{(1)} + E_{\mathrm{dd}}^{(2)} 
\nonumber \\
  & \rightarrow & \lambda^{2+2\alpha}E_{\mathrm{kin}} 
    + \lambda^{-2+\alpha}E_{\mathrm{trap}} \nonumber \\
  & &\mbox{} + \lambda^{3+5\alpha/2}E_\mathrm{dd}^{(1)} + 
  \lambda^{3+2\alpha}E_\mathrm{dd}^{(2)}\, .
\end{eqnarray}

\subsection{Virial theorem}
Since the minimum of $E$ is achieved by the true ground-state density, all
first-order changes of $E$ in the vicinity of ${\lambda=1}$ must be generated by
the explicit change in $N$, ${\delta N = \delta \lambda ~\alpha N}$, so that
\begin{eqnarray}
\alpha N \frac{\partial E}{\partial N} & = & (2+2\alpha)E_\mathrm{kin} +
(-2+\alpha)E_\mathrm{trap} \nonumber \\
  & &
\mbox{} + (3+\tfrac{5}{2}\alpha)E_\mathrm{dd}^{(1)} +(3+2\alpha)E_\mathrm{dd}^{(2)}
\end{eqnarray}
is true for all values of $\alpha$.  Choosing two values of $\alpha$ for
independent statements, we have
\begin{equation}  \label{Eqn:alpha=0}
2E_\mathrm{kin}-2E_\mathrm{trap}
+3\bigl(E_\mathrm{dd}^{(1)}+E_\mathrm{dd}^{(2)}\bigr) = 0
\end{equation}
for $\alpha=0$, and 
\begin{equation} \label{Eqn:alpha=0a}
2E_\mathrm{kin}+10E_\mathrm{trap}+\bigl(E_\mathrm{dd}^{(1)}-E_\mathrm{dd}^{(2)}\bigr) 
= 4N\frac{\partial E}{\partial N}
\end{equation}
for $\alpha=-\frac{4}{3}$, which are supplemented by the first line
of Eq.~(\ref{Eqn:SumE}).  Further, we note the parametric dependence on
$\mu$, $\omega$, and $M$,
\begin{eqnarray}
\mu \frac{\partial}{\partial \mu} E & = & 2E_\mathrm{dd} = 2(E_\mathrm{dd}^{(1)} +
E_\mathrm{dd}^{(2)})\, , \nonumber \\
\omega \frac{\partial}{\partial \omega} E & = & 2E_\mathrm{trap}\, ,  
\nonumber \\
M \frac{\partial}{\partial M} E & = & E_\mathrm{trap} - E_\mathrm{kin}\, .
\label{Eqn:partial}
\end{eqnarray}

Now, owing to the scaling argument, which will be presented next, we find that
$E_\mathrm{dd}^{(1)}/E_\mathrm{dd}^{(2)} \sim \sqrt{N}$, which allows us to neglect
$E_\mathrm{dd}^{(2)}$ for large $N$.  
Applying $E_\mathrm{dd}^{(1)}\pm E_{\mathrm{dd}}^{(2)}\approx E_{\mathrm{dd}}^{(1)}$ 
to the first line of Eq.~(\ref{Eqn:SumE}) and
Eqs.~(\ref{Eqn:alpha=0})--(\ref{Eqn:partial}) then yields
\begin{equation}\label{Eqn:totalEfn}
E(\mu, \omega, M, N) 
\approx \hbar \omega N^{3/2}\mathcal{E}(\epsilon N^{1/4})\,,
\end{equation}
where
\begin{equation}
\epsilon = \frac{\mu_0\mu^2}{4\pi l_0^3} \Big/ (\hbar\omega)
\end{equation}
is a dimensionless interaction strength that can be understood as the ratio
between the interaction energy of two magnetic dipoles separated by
$l_0=\sqrt{\hbar/(M\omega)}$ and the transverse harmonic oscillator energy
scale, and $\mathcal{E}(\ )$ is a dimensionless function of $\epsilon N^{1/4}$.  
We remark that the $N^{3/2}$ dependence in the prefactor results from the
degeneracy of the harmonic confinement in 2D.

\subsection{Dimensionless variables}\label{sSec:DimlessVar}
We define the natural length scale of the system, $a$, the dimensionless
position, $\mathbf{x}$, and the dimensionless density, $g(\mathbf{x})$, 
in accordance with
\begin{equation}
a = l_0 N^{1/4}\,, \quad \mathbf{x} = \frac{\mathbf{r}}{a}\,,\quad 
g(\mathbf{x}) = \frac{a^2}{N}\,n(\mathbf{r})\, ,
\end{equation}
so that the scaled density is normalized to unity.  
Choosing $\hbar\omega N^{3/2}$ as the energy unit, we have
\begin{eqnarray}
\frac{E_\mathrm{TFD}^\mathrm{(2D)}[g]}{\hbar \omega N^{3/2}} & = & \pi \int
\!(d\mathbf{x})\,  g(\mathbf{x})^2 + \frac{1}{2} \int \!(d\mathbf{x})\,
x^2g(\mathbf{x})\nonumber\\
 & & \mbox{}+\epsilon N^{1/4} \biggl( \frac{256}{45} \sqrt{\pi} \int
\!(d\mathbf{x}) \,g(\mathbf{x})^{5/2} \nonumber \\
  & & \mbox{}- N^{-1/2} \pi \!\int \!(d\mathbf{x})
\,g(\mathbf{x}) \sqrt{-\nabla^2} g(\mathbf{x}) \biggr)\,,\qquad
\end{eqnarray}
where $-\nabla^2$ now differentiates with respect to position $\mathbf{x}$, and
the scaled density that minimizes $E_\mathrm{TFD}^\mathrm{(2D)}$ must obey
\begin{eqnarray}\label{Eqn:denEq}
& & 2\pi g(\mathbf{x}) + \frac{1}{2} x^2 + \epsilon N^{1/4} \bigg( \frac{128}{9}
\sqrt{\pi} \,g(\mathbf{x})^{3/2} \nonumber \\
& & \hspace*{6em}\mbox{} - N^{-1/2} 2\pi \sqrt{-\nabla^2}
g(\mathbf{x}) \bigg) = \frac{1}{2} X^2\,,\qquad
\end{eqnarray}
where $\frac{1}{2}X^2$ is the scaled chemical potential.  The term preceded by
$N^{-1/2}$ originates in $E_\mathrm{dd}^{(2)}$, which was neglected for large $N$
on the way to Eq.~(\ref{Eqn:totalEfn}).

\section{Density and energy of the ground state}\label{sec:result}
For $N\sim 10^4$, which is a modest value for typical experiments with
ultracold atoms, the $N^{-1/2}$ term is a correction in the one-percent
regime.  Given that the TFD approximation is generally introducing errors of
the order of a few percent, this term is of a negligible size.  Therefore, we
shall consistently discard it and all other $N^{-1/2}$ terms.  Equation
(\ref{Eqn:denEq}) then reduces to
\begin{equation}\label{Eqn:denEqPT}
\epsilon N^{1/4} \frac{128}{9} \sqrt{\pi} {\sqrt{g(\mathbf{x})}\,}^3 + 2\pi
{\sqrt{g(\mathbf{x})}\,}^2 = \frac{1}{2}(X^2 - x^2)\, ,
\end{equation}
which does not single out any spatial direction and thus implies an isotropic
ground state density, $g(\mathbf{x}) = g(x)$.  We also recognize that
Eq.~(\ref{Eqn:denEqPT}) is a cubic equation for $\sqrt{g(x)}$ and can be
solved analytically.

\begin{figure}
\centerline{\includegraphics{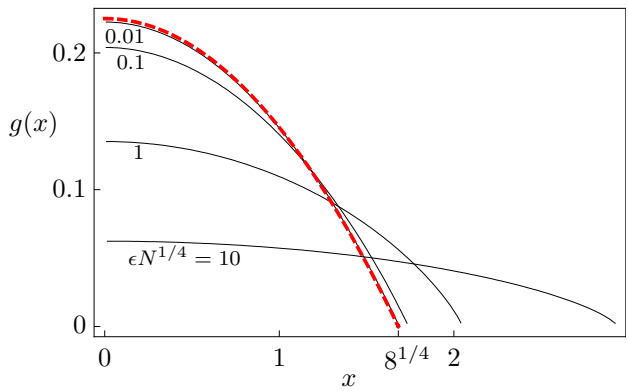}}
\caption{The dimensionless spatial density $g(x)$ at various values of
  $\epsilon N^{1/4} = 0.01, 0.1, 1, 10$ (thin lines). The TF profile
  (thick dashed line) is included as a reference.  Note that there is an
  insignificant difference from the TF profile for ${\epsilon N^{1/4} <
  10^{-2}}$.  }  \label{Fig:DLD}
\end{figure}

In Fig.~\ref{Fig:DLD}, we plot the dimensionless density $g(x)$ for different
values of $\epsilon N^{1/4}$.  We observe that the stronger the dipole
repulsion (larger $\epsilon$), the lower the central density and the larger
the radius of the cloud.  This feature is reminiscent of the condensate wave
function of bosonic atoms when a repulsive contact interaction is taken into
account in the mean-field formalism \cite{dalfovo}.  In contrast to that
exhibited by a 3D spin-polarized Bose-Einstein condensate \cite{pfau}, the
simple symmetry of the isotropic harmonic confinement is \emph{preserved} in
the ground-state density in 2D.  We remark that this is partially a
consequence of choosing the direction of spin polarization along the $z$-axis.
The situation is markedly different, and more interesting, when the
polarization direction breaks the axial symmetry.  This will be discussed in
Sec.~\ref{sec:SDM}.

On the other hand, the dipole interaction for alkali metals are typically
small.  In the limit of $\epsilon \rightarrow 0^+$, we recover the well-known
Thomas-Fermi (TF) profile of noninteracting fermions in a 2D harmonic trap,
\begin{equation}\label{Eqn:TFProf}
n(\mathbf{r}) = \frac{1}{4\pi} l_0^{-4} (R_\mathrm{TF}^2-r^2)
\qquad\mbox{for}\quad 0\leq r\leq R_\mathrm{TF}\,,
\end{equation}
where ${R_\mathrm{TF}=\sqrt{2} (2N)^{1/4} l_0}$ is the Thomas-Fermi radius in
2D.  

\begin{figure}
\centerline{\includegraphics[bb=35 573 300 735,clip=,width=230pt]%
{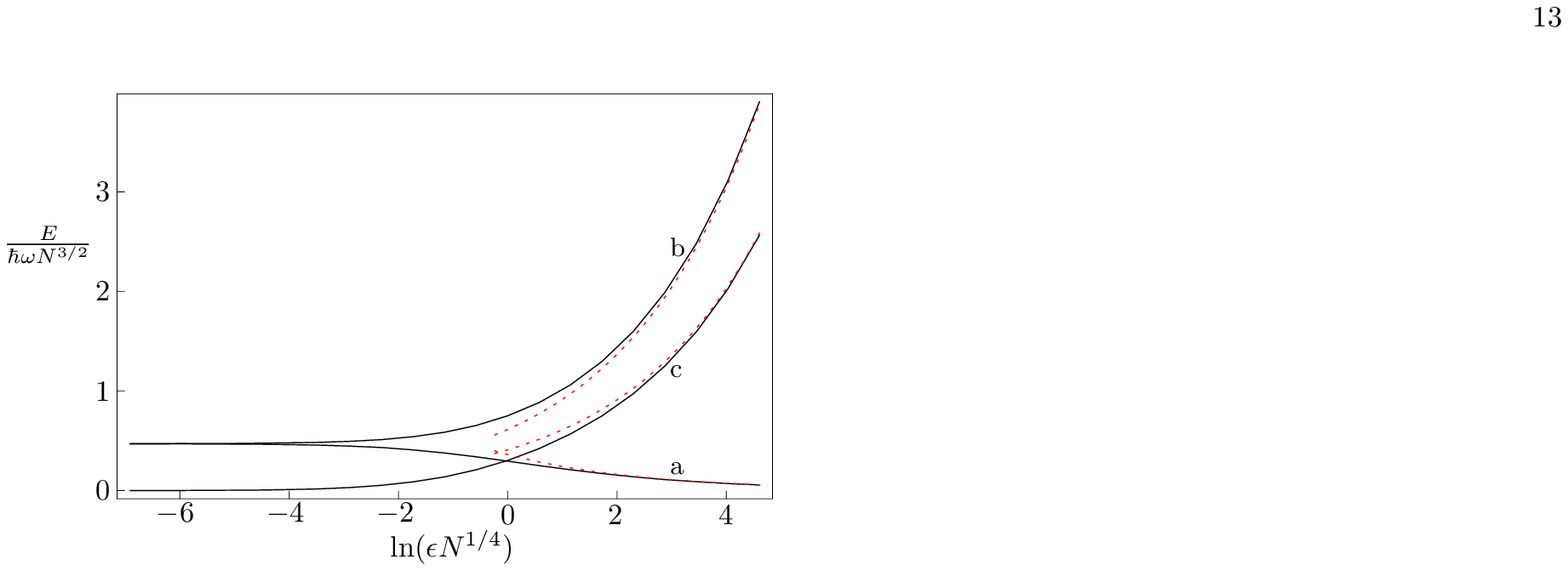}}
\caption{\label{Fig:EvsInt}%
  The energy contributions (a) $E_{\mathrm{kin}}$, (b)
  $E_{\mathrm{trap}}$, and (c) $E_{\mathrm{dd}}^{(1)}$ (in units of 
  ${\hbar\omega N^{3/2}}$) as functions of ${\epsilon N^{1/4}}$ (in logarithmic
  scale).  The solid lines show the full solutions of Eqs.~(\ref{eq:exact-1})
  and (\ref{eq:exact-2}), and exhibit the correct weak-interaction of 
  Eqs.~(\ref{Eqn:smallEpTotalE}) on the far left.  
  The short-dashed lines indicate the asymptotic forms of 
  Eqs.~(\ref{Eqn:Asymptotic}) for large values of $\epsilon N^{1/4}$. 
}
\end{figure}

To evaluate the ground-state energy, we recognize that
Eq.~(\ref{Eqn:denEqPT}) provides a natural way of changing the integration
variable,
\begin{equation}\label{Eqn:changeIntVar}
-x\,dx=(\kappa g^{1/2} + 2 \pi)\, dg 
\quad\mbox{with}\enskip\kappa\equiv\epsilon N^{1/4} \frac{64}{3} \sqrt{\pi}\,,
\end{equation}
where the position dependence of $g$ is left implicit, such that all terms
except $E_\mathrm{dd}^{(2)}$ can be analytically expressed in terms of
$\kappa$ and the central density ${G \equiv g(x=0)}$, 
\begin{eqnarray}\label{eq:exact-1}
  \frac{E_\mathrm{kin}}{\hbar \omega N^{3/2}} 
& = & \frac{4\pi^2}{21}\bigl(3\kappa G^{7/2}+7\pi G^3\bigr)
\,, \nonumber \\[1.2ex] 
\frac{E_\mathrm{trap}}{\hbar \omega N^{3/2}} 
& = & \frac{1}{2}X^2
-\frac{\pi}{21}\bigl(7\kappa^2G^4+40\pi\kappa G^{7/2}+56\pi^2G^3\bigr)
\,, \nonumber \\[1.2ex]
\frac{E_\mathrm{dd}^{(1)}}{\hbar \omega N^{3/2}} 
& = &\frac{2\pi}{105}\bigl(7\kappa^2G^4+16\pi\kappa G^{7/2}\bigr) 
\,.
\end{eqnarray}
The values of $G$ and $X$ are in turn determined by
\begin{eqnarray}\label{eq:exact-2}
\kappa G^{3/2} + 3\pi G & = & 
\frac{3}{4}X^2\,, \nonumber\\
2\kappa G^{5/2} + 5\pi G^2 & = & 
\frac{5}{2\pi}\,,  
\end{eqnarray}
of which the top equation is Eq.~(\ref{Eqn:denEqPT}) for ${x=0}$, and the
bottom equation states the normalization of $g(\mb{x})$ to unit integral.  The
analytic solutions for $E_\mathrm{kin}$, $E_\mathrm{trap}$, and
$E_\mathrm{dd}^{(1)}$ as functions of $\epsilon
N^{1/4}=3\kappa/(64\sqrt{\pi})$ are plotted in Fig.~\ref{Fig:EvsInt}.  On the
far left in the figure, we observe the equipartition of kinetic and trap
energies at vanishing interaction as one expects in the case of a harmonic
trapping potential.

For weakly interacting atoms, i.e.\ $\epsilon N^{1/4} \ll 1$, we obtain the
various contributions to the energy up to the first order in $\epsilon N^{1/4}$,
\begin{eqnarray}
\frac{E_\mathrm{kin}}{\hbar \omega N^{3/2}} 
& = & \frac{\sqrt{2}}{3}-\frac{128}{105\pi} 2^{1/4}\epsilon
N^{1/4}\,, \nonumber \\[1.2ex] 
\frac{E_\mathrm{trap}}{\hbar \omega N^{3/2}} 
& = & \frac{\sqrt{2}}{3}+\frac{128}{105\pi} 2^{1/4}\epsilon
N^{1/4}\,, \nonumber \\[1.2ex]
\frac{E_\mathrm{dd}^{(1)}}{\hbar \omega N^{3/2}} 
& = & \frac{512}{315\pi} 2^{1/4}\epsilon N^{1/4}
\approx 0.615\,\epsilon N^{1/4}\,. \label{Eqn:smallEpTotalE} 
\end{eqnarray}
Note that the sum of $E_{\rm kin}$ and $E_{\rm trap}$ has no first-order
correction.  

The asymptotic values in the limit of large $\epsilon N^{1/4}$ --- shown
as dashed lines in Fig.~\ref{Fig:EvsInt} --- are  given by
\begin{eqnarray}\label{Eqn:Asymptotic}
\frac{E_\mathrm{kin}}{\hbar \omega N^{3/2}} 
& \sim & (\epsilon N^{1/4})^{-2/5}\, , \nonumber \\
\frac{E_\mathrm{trap}}{\hbar \omega N^{3/2}}, \frac{E_\mathrm{dd}^{(1)}}{\hbar
  \omega N^{3/2}} 
& \sim & (\epsilon N^{1/4})^{2/5}\, , \nonumber \\
\frac{E_\mathrm{dd}^{(2)}}{\hbar \omega N^{3/2}} 
& \sim & (\epsilon N^{1/4})^{-1.58}\,,
\end{eqnarray}
where the final power law is obtained by a numerical fit.  
Note that $E_\mathrm{trap}$ and $E_\mathrm{dd}^{(1)}$ have the same 
large-$\epsilon N^{1/4}$ behavior.

\section{Spin-density matrix}\label{sec:SDM}
While the above treatment yields the TFD approximated ground-state density
profile and energy for a 2D cloud of spin-1/2 fermions that are polarized
along the axial direction and are hence repelling each other, the lack of
spherical symmetry of the magnetic-dipole interaction, which is the source of
some interesting predictions \cite{pfau}, is not well reflected due to the
peculiarity of both the configuration and the low dimension.

In order to take the spin-dependent nature of the magnetic-dipole interaction
into consideration, we extend the treatment by (i) introducing an
external magnetic field strong enough to define a local quantization axis; and
(ii) constructing spin-dependent Wigner functions and hence the corresponding
one-body and two-body spin-density matrices.

For an arbitrary time-independent external magnetic field, 
\begin{equation}
\mathbf{B}(\mathbf{r}) = B(\mathbf{r})~\mathbf{e}(\mathbf{r})\,,
\end{equation}
the magnetic energy of a single dipole is given by
\begin{equation}
- \mathbf{B}(\mathbf{r}) \cdot \boldsymbol{\mu} = - B(\mathbf{r})
\mu~\mathbf{e}(\mathbf{r}) \cdot \boldsymbol{\sigma}
\equiv - v(\mathbf{r})~\mathbf{e}(\mathbf{r}) \cdot
\boldsymbol{\sigma}\,.
\end{equation}
The TF-approximated Wigner function is then
\begin{eqnarray}
\underline{\nu}(\mb{r},\mb{p}) & = &  \eta\bigl( -\zeta -\sfrac{p^2}{2M}
-V(\mb{r}) + v(\mb{r})~\mb{e}(\mb{r})\cdot\boldsymbol{\sigma} \bigr)
\nonumber\\
 & = & \frac{1 +\mb{e}(\mb{r})\cdot\bs{\sigma}}{2} \eta \bigl( P_+(\mb{r}) - p
\bigr) \nonumber\\
 & & + \frac{1 -\mb{e}(\mb{r})\cdot\bs{\sigma}}{2} 
\eta \bigl( P_-(\mb{r}) - p\bigr)\,,
\end{eqnarray}
with 
\begin{equation}
P_\pm(\mb{r}) = \bigl[ 2M\bigl( -\zeta -V(\mb{r}) \pm v(\mb{r}) \bigl)
  \bigr]^{1/2}\,,
\end{equation}
and $-\zeta$ is the chemical potential.  The underscore is a reminder that
this Wigner function is $2\times 2$-matrix valued.
As a result, the
single-particle density also has a corresponding spin dependence,
\begin{eqnarray}
\underline{n}(\mb{r}) & = & \frac{1 \!+\mb{e}(\mb{r})\cdot\bs{\sigma}}{2} \pi
\biggl( \frac{P_+(\mb{r})}{2\pi\hbar} \biggr)^2 \nonumber\\[1ex]
 & & \mbox{}+ \frac{1 \!-\mb{e}(\mb{r})\cdot\bs{\sigma}}{2} \pi \biggl(
\frac{P_-(\mb{r})}{2\pi\hbar} \biggr)^2 \nonumber\\[1ex]
 & \equiv & \half \bigl( n(\mb{r}) + s(\mb{r})\; \mb{e}(\mb{r})
\cdot \bs{\sigma}\bigr)\,. 
\end{eqnarray}
We observe that now there are two functions present here, the total density,
$n(\mb{r})$, and the spin-imbalance density, $s(\mb{r})$, which are
constrained by
\begin{equation} \label{Eqn:sConstraint}
|s(\mb{r})|\leq n(\mb{r})\,, 
\end{equation}
but are otherwise independent of each other.  Therefore, the minimization to
achieve the ground-state energy has to be done over both functions under the
constraints of normalization and positivity: Eqs.~(\ref{Eqn:norm2D}) and
(\ref{Eqn:sConstraint}), respectively.  

We can then evaluate the trap, kinetic, and magnetic energy accordingly, 
\begin{eqnarray}
E_\mathrm{trap} & = & \mathrm{tr}_{2\times2}~\int\!(d\mb{r})~\half M\omega^2
r^2 ~\underline{n}(\mb{r}) \nonumber\\
 & = & \int\! (d\mb{r})~ \half M\omega^2r^2~n(\mb{r})\,, \nonumber\\
E_\mathrm{kin} & = & \mathrm{tr}_{2\times2}~\int\!(d\mb{r})
\frac{(d\mb{p})~}{(2\pi\hbar)^2}~
\frac{p^2}{2M}~\underline{\nu}(\mb{r},\mb{p}) \nonumber\\
 & = & \int\! (d\mb{r})~ \frac{\pi\hbar^2}{2M}\bigl[ n(\mb{r})^2 +
 s(\mb{r})^2\bigr],,\nonumber\\
E_\mathrm{mag} & = & -\mathrm{tr}_{2\times 2}~\int\!(d\mb{r})~
\bs{\mu}\cdot\mb{B}(\mb{r})~\underline{n}(\mb{r}) \nonumber\\
 & = &  -\int\!(d\mb{r})~ v(\mb{r}) s(\mb{r})\,.
\end{eqnarray}
To compute the dipole interaction energy, we construct an approximation to the
two-body spin-density matrix
$\underline{\underline{n}}^{(2)}(\mb{r}',\mb{r}'';\mb{r}',\mb{r}'')$ in the
spirit of Eq.~(\ref{eqn:n2Dirac}), starting with the single-particle orbital, 
\begin{equation}
\phi_m(x) = \left( \begin{array}{c}
  \alpha_m(\mb{x})\\ \beta_m(\mb{x}) \end{array} \right),
\end{equation}
where $\alpha_m$ and $\beta_m$ are the spin-up and spin-down components,
$\mb{x}$ is the position variable, while $x$ denotes the combination of the
position and spin variables, so that the ground-state wave function of a $N$
fermion system can be constructed as a Slater determinant,
\begin{equation}
\psi(x_1, \cdots, x_N) 
= \frac{1}{\sqrt{N!}} \det_{m,l}\bigl[ \phi_m(x_l) \bigr]\,.
\end{equation}

\begin{figure}
\centerline{\includegraphics{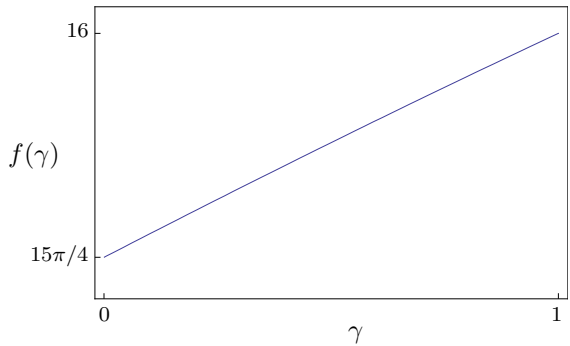}}
\caption{\label{Fig:Elliptic}%
  $f(\gamma)$ at relevant values of $\gamma$.  It can be shown that
  $15\pi/4 < f(\gamma) < 16$ for $0<\gamma<1$.  } 
\end{figure}

\begin{widetext}
When expressing the one-body and two-body spin-density matrices in terms of
single-particle orbitals, we get
\begin{eqnarray}
\underline{n}(\mb{x};\mb{y}) & = & N\int\!dx_2 \cdots dx_N~
\psi(x,x_2,\cdots,x_N) \psi(y,x_2,\cdots,x_N)^* \nonumber\\
 & = & \sum_m \left( \begin{array}{cc} \alpha_m(\mb{x})\alpha_m(\mb{y})^* &
  \alpha_m(\mb{x})\beta_m(\mb{y})^*\\ \beta_m(\mb{x})\alpha_m(\mb{y})^* &
  \beta_m(\mb{x})\beta_m(\mb{y})^* \end{array} \right) 
\equiv \left( \begin{array}{cc} n_\uu(\mb{x};\mb{y}) &
  n_\ud(\mb{x};\mb{y})\\ n_\du(\mb{x};\mb{y}) &
  n_\dd(\mb{x};\mb{y}) \end{array} \right),\nonumber\\[1.2ex]
\underline{\underline{n}}^{(2)}(\mb{x},\mb{y};\mb{x}',\mb{y}') & = &
\frac{N(N-1)}{2}\int\!dx_3 \cdots dx_N~\psi(x,y,x_3,\cdots,x_N)
\psi(x',y',x_3,\cdots,x_N)^* \nonumber\\
 & = & \frac{1}{2} \sum_{l,m} \left( \begin{array}{c}
  \alpha_l(\mb{x})\\ \beta_l(\mb{x}) \end{array} \right) \otimes
\left( \begin{array}{c} \alpha_m(\mb{y})\\ \beta_m(\mb{y}) \end{array} \right)
\left[ \left( \begin{array}{c} \alpha_l(\mb{x}')\\ \beta_l(\mb{x}') \end{array}
  \right) \otimes \left( \begin{array}{c}
    \alpha_m(\mb{y}')\\ \beta_m(\mb{y}') \end{array} \right)
   -\left( \begin{array}{c}
    \alpha_m(\mb{x}')\\ \beta_m(\mb{x}') \end{array} \right) \otimes
  \left( \begin{array}{c} \alpha_l(\mb{y}')\\ \beta_l(\mb{y}') \end{array}
  \right) \right]^\dagger \nonumber\\
 & = & \frac{1}{2} \left(\underline{n}(\mb{x};\mb{x'}) \otimes
\underline{n}(\mb{y};\mb{y'}) -  \big[ \underline{n}(\mb{x};\mb{y'}) \otimes
  \underline{n}(\mb{y};\mb{x'}) \big]_{{\mathrm T}_{23}} \right),
\label{eqn:n2}
\end{eqnarray}
where ${\mathrm T}_{23}$ means interchanging the second and the third columns.
The double summation in $\underline{\underline{n}}^{(2)}$ includes the $l=m$
summands of the self-energy, which has equal contributions to the direct and
exchange terms and hence does not contribute to the sum.

The contact term in the dipole interaction potential is nonvanishing only in
the singlet state,
\begin{eqnarray}
E_\mathrm{dd,s} & = & \mathrm{tr}_{4\times 4}~\int\!(d\mb{r}')(d\mb{r}'')
~\frac{1}{l_z\sqrt{2\pi}} ~ \underline{\underline{n}}^{(2)}
(\mb{r}',\mb{r}'';\mb{r}',\mb{r}'')\,
\frac{\mu_0}{4\pi} \bs{\mu} \cdot \biggl( -\frac{8\pi}{3}
\delta(\mb{r}'-\mb{r}'') \biggr) \bs{\mu} \nonumber\\
 & = & \frac{\sqrt{2\pi}}{l_z} ~\frac{\mu_0\mu^2}{4\pi}  \int\!(d\mb{r})~
\bigl[ n(\mb{r})^2-s(\mb{r})^2 \bigr]\,,
\end{eqnarray}
where the prefactor $\sqrt{2 \pi}/l_z$ originates in the reduction of
dimensionality.  We observe that, owing to the $1/l_z$ scaling, the relative
strength of this term can be tuned by adjusting the stiffness of the
$z$-confining trap.  

On the other hand, the triplet state interacts according to the remaining terms
in the dipole potential.  Since the state is symmetric under particle
exchange, we use, instead of the $\underline{\underline{n}}^{(2)}$ in
Eq.~(\ref{eqn:n2}), an alternative two-body density,
\begin{equation}
\tilde{\underline{\underline{n}}}^{(2)} (\mb{r}',\mb{r}'';\mb{r}',\mb{r}'') =
\frac{1}{2} \Bigl( \underline{n}(\mb{r}')\otimes\underline{n}(\mb{r}'') 
-\underline{n}(\mb{r}';\mb{r}'')\otimes\underline{n}(\mb{r}'';\mb{r}')\Bigr)\,,
\end{equation}
which yields the same energy but greatly simplifies the computation due
to its tensor product structure.  

The triplet interaction energy is then given by
\begin{equation}
E_\mathrm{dd,t} = \mathrm{tr}_{4\times 4} \ \frac{\mu_0\mu^2}{4\pi}
\int\!(d\mb{r}')(d\mb{r}'')~
\tilde{\underline{\underline{n}}}^{(2)} (\mb{r}',\mb{r}'';\mb{r}',\mb{r}'')
\,
\frac{\rho^2\bs{\sigma}\cdot\bs{\tau}- 3\bs{\sigma}\cdot\bs{\rho}
\bs{\rho}\cdot\bs{\tau}}{\rho^5} \,,
\end{equation}
with $\bs{\rho}=\mb{r}'-\mb{r}''$, and $\bs{\tau}$ denotes the Pauli vector
for the second atom.  To evaluate this expression, we apply the same procedure
as that used to obtain Eq.~(\ref{Eqn:Edd2D}) and find
\begin{eqnarray} \label{eqn:65}
E_\mathrm{dd,t}^{(1)} & = & \frac{32}{45}\sqrt{2\pi} \frac{\mu_0 \mu^2}{4\pi}
\int\!(d\mb{r}) \frac{3e_z(\mb{r})^2-1}{2} \biggl[
  \bigl( n(\mb{r}) + s(\mb{r}) \bigr)^{5/2} 
+ \bigl( n(\mb{r}) - s(\mb{r}) \bigr)^{5/2}
  - \frac{f(\gamma)}{8} \bigl( n(\mb{r}) + s(\mb{r}) \bigr)^{3/2} \bigl(
  n(\mb{r}) - s(\mb{r}) \bigr)  \biggr]\,,\nonumber\\
E_\mathrm{dd,t}^{(2)} & = & -\frac{1}{2} \frac{\mu_0 \mu^2}{4\pi}
 \int\! (d\mb{r})(d\mb{r}')~
\frac{\nabla s_z(\mb{r}) \cdot\nabla' s_z(\mb{r}') - \nabla\cdot\mb{s}(\mb{r})
  \nabla'\cdot\mb{s}(\mb{r}')}{|\mb{r}-\mb{r}'|}\,,
\end{eqnarray}
where $\gamma = \bigl[P_-(\mb{r})/P_+(\mb{r})\bigr]^2$ is essentially the ratio
between the Fermi energies of the minority and majority spin components, and
\begin{equation}
f(\gamma) = (\gamma^{-1} + 14 + \gamma) E(\gamma) 
+ (-\gamma^{-1} - 6 +7\gamma) K(\gamma)
\end{equation}
is a combination of elliptic integrals that is smooth and finite for
$0<\gamma<1$.  It is clear from Fig.~\ref{Fig:Elliptic} that $f(\gamma)$ can
be replaced by a linear function $\tilde{f}(\gamma) = \frac{15}{4}\pi +
(16-\frac{15}{4}\pi) \gamma$ to simplify computations.  

In passing, we note that a magnetic field with a large component in the
$xy$-plane, such that $3e_z(\mb{r})^2-1<0$ in Eq.\ (\ref{eqn:65}) for some
region, may lead to an energy that is not bounded from below.  The system is
then instable and will collapse and explode within miliseconds (this
catastrophe was observed in dipolar bose gases \cite{pfau2,pfau3}).  When this
happens, so much energy is made available that the 2D confinement will be
lost.

For the simple case of a constant external magnetic field, we have
\begin{equation}
\mb{B}(\mb{r}) = B_0 \mb{e_z}\,,\qquad
v(\mb{r}) = B_0 \mu \equiv v_0\,.
\end{equation}
In the dimensionless quantities, we re-parameterize the spin-imbalance density
in accordance with
\begin{equation}
h(\mb{x}) = \frac{a^2}{N} s(\mb{r}) \equiv \cos
\big(\vartheta(\mb{x})\big)~g(\mb{x})\,,
\end{equation}
so that Eq.~(\ref{Eqn:sConstraint}) is automatically fulfilled.  In the limit
of weak interaction, we obtain the total energy as a functional of both
$g(\mb{x})$ and $\vartheta(\mb{x})$,
\begin{equation}
\frac{E_\mathrm{TFD}^\mathrm{(2D)}[g,\vartheta]}{\hbar \omega N^{3/2}} 
= \frac{\pi}{2} \int
\!(d\mathbf{x})\,  g(\mathbf{x})^2 
\biggl(1 + \cos^2\!\big(\vartheta(\mb{x})\big)\biggr)
+ \frac{1}{2} \int \!(d\mathbf{x})\, x^2g(\mathbf{x})
- \frac{1}{\sqrt{N}}\frac{v_0}{\hbar\omega} \int \!(d\mathbf{x})
\,g(\mathbf{x}) \cos\bigl(\vartheta(\mb{x})\bigr)\,,
\end{equation}
and the variation of $g(\mb{x})$ and $\vartheta(\mb{x})$, with $\half X^2$ as
the Lagrange multiplier, yields
\begin{eqnarray}
\pi g(\mb{x}) + \cos\bigl(\vartheta(\mb{x})\bigr) \biggl( \pi g(\mb{x})
\cos\bigl(\vartheta(\mb{x}) \bigr) - \frac{1}{\sqrt{N}} \frac{v_0}{\hbar\omega}
\biggr) & = & \frac{1}{2} (X^2-x^2)\,,\nonumber\\
g(\mb{x}) \sin\bigl( \vartheta(\mb{x}) \bigr) \biggl( \pi g(\mb{x})
\cos\bigl(\vartheta(\mb{x}) \bigr) 
- \frac{1}{\sqrt{N}} \frac{v_0}{\hbar\omega}\biggr) & = & 0\,. 
\end{eqnarray}
\end{widetext}

\begin{figure}
\centerline{\includegraphics[width=230pt]{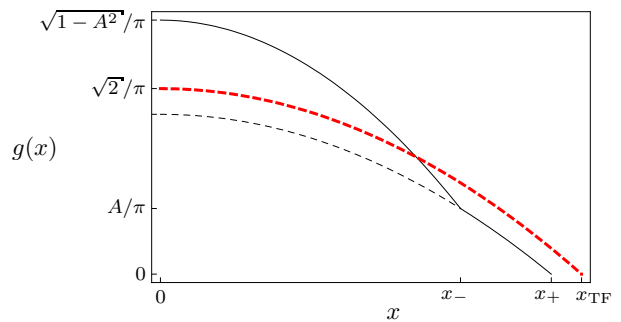}}
\caption{\label{Fig:SM}%
  The dimensionless density profile of a SM (thin solid line) with
  $A=0.25$, in comparison with the TF profile (i.e.\ SP solution, thick dashed
  line).  The thin dashed line indicates the density of the majority component
  in the spin mixture.  For greater values of $A$ while keeping $A\leq
  1/\sqrt{2}$, the density profile approaches that of the SP solution with a
  lowering central density and an increasing $x_\pm$.  In the opposite limit,
  we recover a mixture of equal spin-components when there is no external
  magnetic field, i.e.\ $A=0$.  } 
\end{figure}

There are two nontrivial solutions, one for a spin-polarized (SP) system and
the other allowing a spin-mixture (SM).  The SP solution yields the TF profile
equivalent to Eq.~(\ref{Eqn:TFProf}).  The SM solution gives
\begin{equation}
g(\mb{x}) = \left\{ 
\begin{array}{ll}
\displaystyle\frac{1}{2\pi} \bigl(2\sqrt{1-A^2}-x^2\bigr) 
& \mbox{for}\ 0\leq x\leq x_-\,,\\[2ex]
\displaystyle\frac{1}{2\pi} \biggl( A+\sqrt{1-A^2}-\frac{x^2}{2} \biggr) 
&\mbox{for}\ x_-\leq x\leq x_+\,,
\end{array}\right.
\end{equation}
with 
\begin{equation}
A = \frac{v_0}{\hbar\omega} \frac{1}{\sqrt{N}} = \frac{B_0 \mu}{\pi
  \sqrt{N}},
\end{equation}
together with a constant spin-imbalance density in the center,
\begin{equation}
g(\mb{x})~\cos\!\big(\vartheta(\mb{x})\big) = \frac{A}{\pi}, 
\end{equation}
where $x_\pm$ are the radii of the spin-mixture (lower sign) and the entire
cloud (upper sign) respectively, given by
\begin{equation}
x_\pm = 2\bigl( \sqrt{1-A^2}\pm A \bigr)\,.
\end{equation}  
These matters are illustrated in Fig.~\ref{Fig:SM}.

However, the existence of a spin-mixture requires an extremely weak external
magnetic field, such that
\begin{equation}
\frac{v_0}{\hbar\omega}\frac{1}{\sqrt{N}} \leq \frac{1}{\sqrt{2}}\,.
\end{equation}
This condition arises from the positivity of the radii $x_\pm$.  
In usual experimental set-ups, this translates into 
${B\sim 10 \,\mathrm{mG}}$ for a system of ${N=10^6}$ atoms with a radial
harmonic confinement of ${\omega = 2\pi\times20\,\mathrm{Hz}}$.  
In other words, a spin-polarized cloud is readily attainable.
This justifies the treatment of a spin-polarized system before
Sec.\ \ref{sec:SDM}.

\section{Summary and Outlook}\label{sec:SaO}

\begin{table*}
\caption{\label{Tab:123D}Summary of the density functionals for the kinetic
  energy and the dipole-dipole interaction energy in one, two, and three
  dimensions.  In 1D, the spins are polarized normal to the $z$-axis, along
  which the atoms align.  $\mathrm{Erfc}(~)$ denotes the complementary error
  function, and $t=|z-z'|/(\sqrt{2}l_\perp)$.  Note that further
  simplification of the 1D expression of $E_\mathrm{dd}$ involves taking the
  limit of $l_\perp \rightarrow 0^+$ which should be done with extreme care.
  In 2D and 3D, the spins are polarized along the $z$-direction, $\theta$ in
  3D refers to the azimuthal angle of the vector $\mb{r}-\mb{r}'$.  }
\centering
\begin{ruledtabular}
\begin{tabular}{@{\hspace{1em}}cll@{\hspace{1em}}}
 & $E_\mathrm{kin}$ & $E_\mathrm{dd}$ \\[1ex]
\hline
\rule[7pt]{0pt}{9pt}1D & \dis{\int\!dz~\frac{\pi^2 \hbar^2}{6M} n(z)^3} &
  \dis{\frac{\mu_0\mu^2}{4\pi} \int\!dz~dz' \biggl( n(z)
    n(z') - n(z;z') n(z';z) \biggr) \frac{1}{\sqrt{2}l_\perp^3} 
\left[ \Bigl(\frac{1}{2}
      + t^2 \Bigr)\sqrt{\pi}e^{t^2} \mathrm{Erfc}(t) - t \right] } \\[2.8ex]
2D & \dis{\int\!(d\mb{r}_\perp)~\frac{\pi\hbar^2}{2M} n(\mb{r}_\perp)^2} &
\dis{\frac{\mu_0\mu^2}{4\pi}\!\int\!(d\mb{r}_\perp)
\biggl[ \frac{256}{45}\sqrt{\pi}n(\mb{r}_\perp)^{5/2} 
- \pi n(\mb{r}) \sqrt{-\nabla^2} n(\mb{r}_\perp) \biggr]}
\\[2.2ex]
3D & \dis{\int\!(d\mb{r})~\frac{\hbar^2}{20\pi^2 M} \bigl[ 6\pi^2
    n(\mb{r})\bigr]^{5/3}} & \dis{\frac{\mu_0\mu^2}{4\pi}
  \int\!(d\mb{r})(d\mb{r}')~ \frac{1}{2}~n(\mb{r})~
  \frac{1-3\cos^2\theta}{|\mb{r}-\mb{r}'|^3} ~n(\mb{r}')}\rule[-10pt]{0pt}{2pt}
\end{tabular}
\end{ruledtabular}
\end{table*}

Table \ref{Tab:123D} summarizes the kinetic and the dipole-dipole interaction
energies as functionals of the single-particle density for a fully
spin-polarized gas in one, two, and three dimensions.  It is clear that the
structure of the density functionals depends crucially on the spatial dimension.
The procedure used here to reduce dimensionality is by no means unique, but
fairly well justified by the strong confinement of a stiff harmonic trap in a
possible experimental set-up.

In 2D, the total energy with TFD approximation depends on both the
dimensionless interaction strength $\epsilon$, and the number of particles
$N$, as it does in 3D, but the $N$-dependence is slightly more complicated.
Namely, as one piece of the interaction energy is proportional to $\epsilon
N^{1/4}$, while the other piece is proportional to $\epsilon N^{-1/4}$, the
latter is always a factor of $\sqrt{N}$ smaller, inviting a perturbative
treatment.

For large $\epsilon N^{1/4}$, the potential energy and the first piece of the
interaction energy dominate, and are of the order of 
$\hbar\omega N^{3/2}\times(\epsilon N^{1/4})^{2/5}$,
while the kinetic energy is of the order of
$\hbar\omega N^{3/2}\times(\epsilon N^{1/4})^{-2/5}$. 
Numerical results suggest that the second piece of the
interaction energy is of the order 
$\hbar\omega N^{3/2}\times(\epsilon N^{1/4})^{-1.58}$, which makes
it the most slowly growing term in the total energy.

In the limiting case of $\epsilon \rightarrow 0^+$, the kinetic and potential
energies are both $\hbar \omega N^{3/2} \frac{\sqrt{2}}{3}$, where the
equality is well predicted by the virial theorem applied to a simple harmonic
oscillator.  The two pieces of the interaction energy are of the order
$N^{7/4}$ and $N^{5/4}$ respectively, even though both are vanishing due to
small $\epsilon$.

In addition to the fully spin-polarized situation, we also dealt with
partially polarized gases, allowing for inhomogeneous polarization.  By
considering the spin-density matrix, we found the energy as a functional of
the total density and the spin-imbalance density, and then determined the
implied ground-state density profile.  
For typical experimental parameters and a modest number of trapped
atoms, a spin-mixture can only exist for an extremely weak external magnetic
field.  
In other words, a fully spin-polarized gas is readily attainable.

Having thus established the TFD functionals, we intend to investigate the
excitation energies of the system for small deviation from the equilibrium.
On the other hand, it is well-known that the TF approximation is problematic
at the boundary of the system.  We will follow up on the gradient corrections
of von Weizs\"acker type.  It is perceivable that once the corrections are
included, $E_\mathrm{dd}^{(2)}$ may no longer be negligible.  Lastly, we would
like to explore other external trapping potentials, such as anisotropic
harmonic traps, possibly with an optical lattice superimposed.

\acknowledgments
We are grateful for discussions with Kazimierz Rz\c{a}\.{z}ewski. Centre for
Quantum Technologies is a Research Centre of Excellence funded by Ministry of
Education and National Research Foundation of Singapore.

\begin{widetext}
\appendix*\renewcommand{\theequation}{A\arabic{equation}}
\section{Calculating the interaction energy}

The splitting of the dipole-dipole interaction energy into direct and exchange
energy at the level of the one-particle density and the one-particle density
matrix turns out to be inconvenient at it stands, since this yields two
integrals, which both diverge individually but together sum up to a finite
value.  This prompts us to express everything in terms of the one-particle
Wigner function,
\begin{eqnarray}
E_\mathrm{dd} & = & \frac{1}{2} \frac{\mu_0\mu^2}{4\pi} \!\int \!
\frac{(d\mathbf{r}')(d\mathbf{r}'')(d\mathbf{p}_1)(d\mathbf{p}_2)}
{(2\pi\hbar)^4} \,\frac{1}{|\mathbf{r}'-\mathbf{r}''|^3}
\bigg[ \nu(\mathbf{r}',\mathbf{p}_1) \,\nu(\mathbf{r}'',\mathbf{p}_2) 
- \nu {\left( \mbox{$\frac{\mathbf{r}'+\mathbf{r}''}{2}$}, \mathbf{p}_1 
\right)} \nu
{\left( \mbox{$\frac{\mathbf{r}'+\mathbf{r}''}{2}$}, \mathbf{p}_2 \right)} 
e^{i(\mathbf{p}_1-\mathbf{p}_2) \cdot (\mathbf{r}'-\mathbf{r}'')/\hbar}
\bigg] \nonumber \\
& = & \frac{1}{2} \frac{\mu_0\mu^2}{4\pi} \!\int \!
\frac{(d\mathbf{r})(d\boldsymbol{\rho})(d\mathbf{p}_1)(d\mathbf{p}_2)}
{(2\pi\hbar)^4}(d\mathbf{k}_1) (d\mathbf{k}_2)
\frac{1}{\rho^3} \,e^{i (\mathbf{k}_1+\mathbf{k}_2) \cdot \mathbf{r}}
\overline{\nu}(\mathbf{k}_1,\mathbf{p}_1)
\overline{\nu}(\mathbf{k}_2,\mathbf{p}_2) \!\left(
  e^{i(\mathbf{k}_1-\mathbf{k}_2) \cdot \boldsymbol{\rho}/2} -
  e^{i(\mathbf{p}_1-\mathbf{p}_2) \cdot \boldsymbol{\rho}/\hbar} \right),
\end{eqnarray}
\end{widetext}\renewcommand{\theequation}{A\arabic{equation}}
where we have used the substitution ${\mathbf{r} = \half
(\mathbf{r}'+\mathbf{r}'')}$, ${\boldsymbol{\rho} =
\mathbf{r}'-\mathbf{r}''}$ and the Fourier transform in 2D,
\begin{eqnarray}
\nu(\mathbf{r},\mathbf{p}) & = & \int \!(d\mathbf{k})
\,\overline{\nu}(\mathbf{k},\mathbf{p}) e^{i\mathbf{k}\cdot\mathbf{r}}\, ,
\nonumber \\
\overline{\nu}(\mathbf{k},\mathbf{p}) 
& = & \int \!\frac{(d\mathbf{r})}{(2\pi)^2}
\,\nu(\mathbf{r},\mathbf{p}) e^{-i\mathbf{k}\cdot\mathbf{r}}\, .  
\end{eqnarray}
The integration over ${\bs{\rho}}$ can be evaluated with the outcome
\begin{eqnarray}\label{Eqn:intRho}
 & & \int \!(d\boldsymbol{\rho}) \,\frac{1}{\rho^3} \left(
   e^{i(\mathbf{k}_1-\mathbf{k}_2) \cdot \boldsymbol{\rho}/2} -
   e^{i(\mathbf{p}_1-\mathbf{p}_2) \cdot \boldsymbol{\rho}/\hbar} \right)
 \nonumber \\
 & = & 2\pi \bigl( |\mathbf{p}_1-\mathbf{p}_2|/\hbar -
   |\mathbf{k}_1-\mathbf{k}_2|/2 \bigr)\,.
\end{eqnarray}
We recognize that the interaction energy is split into two pieces, 
\begin{eqnarray}
  E_\mathrm{dd} & = & E_\mathrm{dd}^{(1)} + E_\mathrm{dd}^{(2)}\, , \nonumber \\
  E_\mathrm{dd}^{(1)} & \equiv & \frac{1}{2} \frac{\mu_0\mu^2}{4\pi} 2\pi \int
  (d\mathbf{r}) \frac{(d\mathbf{p}_1)}{(2\pi\hbar)^2}
  \frac{(d\mathbf{p}_2)}{(2\pi\hbar)^2} (d\mathbf{k}_1) (d\mathbf{k}_2)
  \nonumber\\
  & & \times e^{i (\mathbf{k}_1+\mathbf{k}_2) \cdot \mathbf{r}}
  \,\overline{\nu}(\mathbf{k}_1,\mathbf{p}_1)
  \,\overline{\nu}(\mathbf{k}_2,\mathbf{p}_2)
  \frac{|\mathbf{p}_1-\mathbf{p}_2|}{\hbar} \,, \nonumber \\
  E_\mathrm{dd}^{(2)} & \equiv & -\,\frac{1}{2} \frac{\mu_0\mu^2}{4\pi} 2\pi \int
  (d\mathbf{r}) \frac{(d\mathbf{p}_1)}{(2\pi\hbar)^2}
  \frac{(d\mathbf{p}_2)}{(2\pi\hbar)^2} (d\mathbf{k}_1) (d\mathbf{k}_2)
  \nonumber\\
  & & \times e^{i (\mathbf{k}_1+\mathbf{k}_2) \cdot \mathbf{r}}
  \,\overline{\nu}(\mathbf{k}_1,\mathbf{p}_1)
  \,\overline{\nu}(\mathbf{k}_2,\mathbf{p}_2)
  \frac{|\mathbf{k}_1-\mathbf{k}_2|}{2} \,,  \qquad
\end{eqnarray}
but this is \emph{not} the splitting into the direct and exchange
terms, as the integration of a single exponential term in
Eq.~(\ref{Eqn:intRho}) will not converge.

A closer look at $E_\mathrm{dd}^{(1)}$ tells us that the $\int (d\mathbf{k}_1)$
and $\int (d\mathbf{k}_2)$ integrations recover the Wigner functions, which
impose an upper limit of $P=\hbar\sqrt{4\pi n(\mathbf{r})}$
on the length of $p_1$ and $p_2$, so that
\begin{eqnarray}
  E_\mathrm{dd}^{(1)} & \equiv & \frac{1}{2} \frac{\mu_0\mu^2}{4\pi}
  2\pi \int (d\mathbf{r}) \frac{(d\mathbf{p}_1)}{(2\pi\hbar)^2}
  \frac{(d\mathbf{p}_2)}{(2\pi\hbar)^2} |\mathbf{p}_1-\mathbf{p}_2|/\hbar
  \nonumber \\ 
  & = & \frac{\mu_0\mu^2}{4\pi} \int \!(d\mathbf{r})
  \frac{256}{45} \sqrt{\pi}\, n(\mathbf{r})^{5/2}\, .
\end{eqnarray}
On the other hand, $E_\mathrm{dd}^{(2)}$ needs to be treated differently.  The
$\int (d\mathbf{p}_1)$ and $\int (d\mathbf{p}_2)$ integrations yield the form
factor 
\begin{equation}
\overline{n}(\mathbf{k}) = \int \!\frac{(d\mathbf{p})}{(2\pi\hbar)^2}
\,\overline{\nu}(\mathbf{k},\mathbf{p}) = \int \!\frac{(d\mathbf{r})}{(2\pi)^2}
\,e^{-i\mathbf{k}\cdot\mathbf{r}} n(\mathbf{r})\, ,
\end{equation}
while the $\int (d\mathbf{r})$ integration gives rise to a 2D Dirac delta
function,
\begin{eqnarray}
\int \!(d\mathbf{r}) \,e^{i(\mathbf{k}_1+\mathbf{k}_2)\cdot\mathbf{r}} = (2\pi)^2
\,\delta(\mathbf{k}_1+\mathbf{k}_2)\,.\\\nonumber
\end{eqnarray}
This then takes care of one of the integrations over $\mb{k}_1$ or $\mb{k}_2$,
and we arrive at
\begin{eqnarray}
E_\mathrm{dd}^{(2)} & = & -\frac{1}{2} \frac{\mu_0\mu^2}{4\pi} (2\pi)^3 \int
\!(d\mathbf{k}) \,\overline{n}(\mathbf{k}) \,k \,\overline{n}(-\mathbf{k})
\nonumber \\
 & = & -\frac{\mu_0\mu^2}{4\pi} \pi \int \!(d\mathbf{r}) \,n(\mathbf{r})
\sqrt{-\nabla^2} n(\mathbf{r})\, ,
\end{eqnarray}
where the short-hand notation of $\sqrt{-\nabla^2}$ for the integral operator
(\ref{Eqn:RNL}) is used in recognition that it is equivalent to the integral
operator $-\nabla^2$ when applied twice.  This completes the derivation of
Eqs.~(\ref{Eqn:Edd2D}).

\end{document}